\documentclass[twocolumn,superscriptaddress,floatfix,preprintnumbers]{revtex4}
\usepackage{amsmath, amsfonts}    
\usepackage{graphicx}   
\usepackage{verbatim}   
\usepackage{color}      
\usepackage{subfigure}  
\usepackage{hyperref} 
\usepackage{natbib}
\usepackage{siunitx}
\usepackage{physics}
\usepackage{ulem}
\usepackage[percent]{overpic}
\usepackage{booktabs}
\usepackage{float}
\graphicspath{{./Figure/}}

\usepackage{xspace}
\usepackage[dvipsnames]{xcolor}

\usepackage[acronym]{glossaries}
\makeglossaries

\newcommand{\ignore}[1]{ }

\newcommand{\hwater}{H$_2$O\xspace}
\newcommand{\dwater}{D$_2$O\xspace}

\newcommand{\gH}{\gamma_{\text{H}} }

\newcommand{\wl}{\omega_{\text{L}} }

\newacronym{AFM}{AFM}{atomic force microscopy} 
\newcommand{\afm}{\gls*{AFM}\xspace}

\newacronym{AMO}{AMO}{atomic, molecular, and optical physics}

\newacronym{COSY}{COSY}{correlated spectroscopy} 

\newacronym{CPMG}{CPMG}{Carr-Purcell-Meiboom-Gill} 

\newacronym{CVD}{CVD}{chemical vapor deposition}

\newacronym{DD}{DD}{dynamical decoupling}

\newacronym{ESR}{ESR}{electron spin resonance} 
\newcommand{\esr}{\gls*{ESR}\xspace}

\newacronym{DEER}{DEER}{double electron–electron resonance} 
\newcommand{\deer}{\gls*{DEER}\xspace}

\newacronym{FID}{FID}{free induced decay}

\newacronym{LDW}{LDW}{low-dimensional water} 
\newcommand{\ldw}{\gls*{LDW}\xspace}

\newacronym{MFM}{MFM}{magnetic force microscopy}

\newacronym{MRFM}{MRFM}{magnetic resonance force microscopy}

\newacronym{MRI}{MRI}{magnetic resonance imaging}

\newacronym{MW}{MW}{microwave}

\newacronym{NMR}{NMR}{nuclear magnetic resonance} 
\newcommand{\nmr}{\gls*{NMR}\xspace}

\newacronym{NQR}{NQR}{nuclear quadrupole resonance}

\newacronym{NV}{NV}{nitrogen-vacancy} 
\newcommand{\NV}{\Gls*{NV}\xspace}
\newcommand{\nv}{\gls*{NV}\xspace}

\newacronym{ODMR}{ODMR}{optically detected magnetic resonance} 
\newcommand{\odmr}{\gls*{ODMR}\xspace}

\newacronym{Qdyne}{Qdyne}{quantum heterodyne}

\newacronym{RIE}{RIE}{eactive ion etching}

\newacronym{RF}{RF}{radio frequency}

\newacronym{SEM}{SEM}{scanning electron microscopy}

\newacronym{STM}{STM}{scanning tunneling microscopy} 
\newcommand{\stm}{\gls*{STM}\xspace}

\newacronym{TEM}{TEM}{transmission electron microscopy} 
\newcommand{\tem}{\gls*{TEM}\xspace}

\newacronym{UDD}{UDD}{Uhrig dynamical decoupling}

\newacronym{UHV}{UHV}{ultra-high vacuum}

\begin{document}

\title{
Discovery of an anomalous non-evaporating sub-nanometre water layer in open environment
} 

\author{Zhijie Li}
\thanks{These authors contributed equally to this work.}
\affiliation{CAS Key Laboratory of Microscale Magnetic Resonance and School of Physical Sciences, University of Science and Technology of China, Hefei 230026, China}
\affiliation{Anhui Province Key Laboratory of Scientific Instrument Development and Application, University of Science and Technology of China, Hefei 230026, China}

\author{Xi Kong}
\thanks{These authors contributed equally to this work.}
\affiliation{The State Key Laboratory of Solid State Microstructures and Department of Physics, Nanjing University, 210093 Nanjing, China}

\author{Haoyu Sun}
\affiliation{CAS Key Laboratory of Microscale Magnetic Resonance and School of Physical Sciences, University of Science and Technology of China, Hefei 230026, China}
\affiliation{Anhui Province Key Laboratory of Scientific Instrument Development and Application, University of Science and Technology of China, Hefei 230026, China}

\author{Guanyu Qu}
\affiliation{CAS Key Laboratory of Microscale Magnetic Resonance and School of Physical Sciences, University of Science and Technology of China, Hefei 230026, China}
\affiliation{Anhui Province Key Laboratory of Scientific Instrument Development and Application, University of Science and Technology of China, Hefei 230026, China}

\author{Pei Yu}
\affiliation{CAS Key Laboratory of Microscale Magnetic Resonance and School of Physical Sciences, University of Science and Technology of China, Hefei 230026, China}
\affiliation{Anhui Province Key Laboratory of Scientific Instrument Development and Application, University of Science and Technology of China, Hefei 230026, China}

\author{Tianyu Xie}
\affiliation{CAS Key Laboratory of Microscale Magnetic Resonance and School of Physical Sciences, University of Science and Technology of China, Hefei 230026, China}
\affiliation{Anhui Province Key Laboratory of Scientific Instrument Development and Application, University of Science and Technology of China, Hefei 230026, China}

\author{Zhiyuan Zhao}
\affiliation{CAS Key Laboratory of Microscale Magnetic Resonance and School of Physical Sciences, University of Science and Technology of China, Hefei 230026, China}
\affiliation{Anhui Province Key Laboratory of Scientific Instrument Development and Application, University of Science and Technology of China, Hefei 230026, China}

\author{Guoshen Shi}
\affiliation{Shanghai Applied Radiation Institute, State Key Laboratory Advanced Special Steel, Shanghai University, Shanghai 200444, China}

\author{Ya Wang}
\affiliation{CAS Key Laboratory of Microscale Magnetic Resonance and School of Physical Sciences, University of Science and Technology of China, Hefei 230026, China}
\affiliation{Anhui Province Key Laboratory of Scientific Instrument Development and Application, University of Science and Technology of China, Hefei 230026, China}
\affiliation{Hefei National Laboratory, University of Science and Technology of China, Hefei 230088, China}

\author{Fazhan Shi}
\email{fzshi@ustc.edu.cn}
\affiliation{CAS Key Laboratory of Microscale Magnetic Resonance and School of Physical Sciences, University of Science and Technology of China, Hefei 230026, China}
\affiliation{Anhui Province Key Laboratory of Scientific Instrument Development and Application, University of Science and Technology of China, Hefei 230026, China}
\affiliation{Hefei National Laboratory, University of Science and Technology of China, Hefei 230088, China}
\affiliation{School of Biomedical Engineering and Suzhou Institute for Advanced Research, University of Science and Technology of China, Suzhou 215123, China}

\author{Jiangfeng Du}
\email{djf@ustc.edu.cn}
\affiliation{CAS Key Laboratory of Microscale Magnetic Resonance and School of Physical Sciences, University of Science and Technology of China, Hefei 230026, China}
\affiliation{Anhui Province Key Laboratory of Scientific Instrument Development and Application, University of Science and Technology of China, Hefei 230026, China}
\affiliation{Hefei National Laboratory, University of Science and Technology of China, Hefei 230088, China}
\affiliation{Institute of Quantum Sensing and School of Physics, Zhejiang University, Hangzhou 310027, China}

\begin{abstract}
Water exhibits complex behaviors as a result of hydrogen bonding, and low-dimensional confined water plays a key role in material science, geology, and biology science. Conventional techniques like STM, TEM, and AFM enable atomic-scale observations but face limitations under ambient conditions and surface topographies. NV center magnetic resonance technology provides an opportunity to overcome these limitations, offering non-contact atomic-scale measurements with chemical resolution capability. 
In this study, a nanoscale layer dissection method was developed utilizing NV center technology to analyze water layers with diverse physicochemical properties.
It unveiled the presence of a non-evaporating sub-nanometer water layer on a diamond surface under ambient conditions. This layer demonstrated impervious to atmospheric water vapor and exhibited unique electronic transport mediated via hydrogen bonding.
These findings provide new perspectives and a platform for studying the structure and behavior of low-dimensional water, as well as the surface properties influenced by adsorbed water under native conditions.
\end{abstract}

\maketitle

Water, as the most valuable natural resource, intricately influences life sciences, sculpting the Earth's landscape and captivating the scientific community's interest. Despite its simple chemical structure, the presence of hydrogen (H) bonds leads to numerous phases; currently, there are 17 experimentally confirmed ice polymorphs and several more predicted computationally \cite{salzmann2019Advances}. And among them, low-dimensional adsorbed water or confined water often plays a pivotal role, such as the importance of confined water in the structural and dynamic functions of molecular cell biology \cite{ball2008Water,tanford1962Contribution}. However, due to the influences of restricted dimensions, polarities of restricted surfaces, static electricity, hydrophobic forces, H bonding, etc. \cite{wu2017Wettability, davies2024How}, low-dimensional water exhibits surprisingly rich and diverse phase behaviors \cite{kapil2022Firstprinciples, davies2024How}. Hence, investigating and characterizing the structure and properties of low-dimensional interface water have emerged as crucial research areas.

Low-dimensional water has been extensively researched under various conditions utilizing techniques such as \stm \cite{nie2010Pentagons,maier2016Growth,lin2018TwoDimensional}, \tem \cite{girit2009Graphene,algara-siller2015Square}, and \afm \cite{peng2018Effect,shiotari2017Ultrahighresolution,ma2020Atomic}. These methods facilitate the direct observation of the structural characteristics of low-dimensional water at an atomic scale. However, challenges arise due to interactions involving voltage, charge, and van der Waals forces, making it difficult to entirely avoid impacting the low-dimensional water layers. Another critical issue is the complexity of achieving spatial resolution while analyzing the chemical environment in terms of atomic species and scale \cite{du2024Singlemolecule,claridge2011Electrons}, hindering the direct correlation of other properties and structural features of low-dimensional water. Fortunately, the emerging magnetic resonance technique based on \nv center in diamond presents a unique opportunity. \NV center enable non-contact atomic-scale measurements utilizing magnetic resonance technique, providing atomic spatial resolution without interfering with the sample itself \cite{wang2019Nanoscale,grinolds2014Subnanometre}. Most importantly, NV centers offer chemical shift resolution \cite{abobeih2019Atomicscale,aslam2017Nanoscale} in nuclear magnetic resonance spectroscopy, delivering crucial information for property detection.

\begin{figure}[ht]\centering
  \begin{overpic}[width=0.47\textwidth]{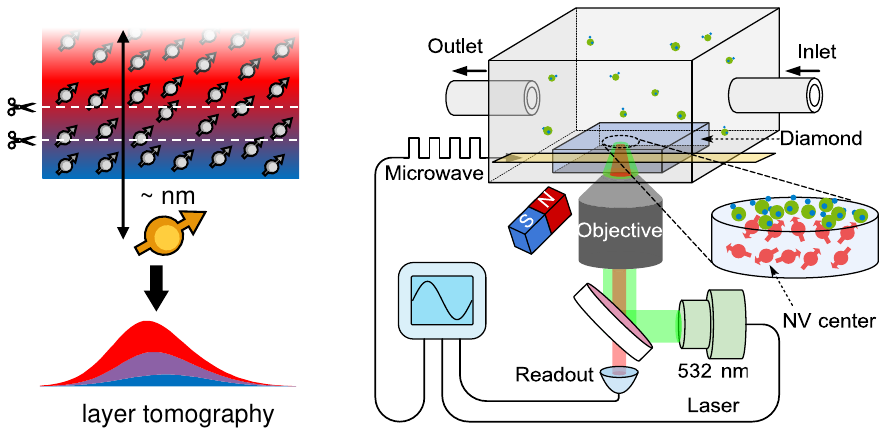} 
     \put (0, 47) {\bf a}
     \put (40, 47) {\bf b}
  \end{overpic}
  \caption[NV Center]{(a) Dissection of the adsorbed water layer on the diamond surface. By manipulating the ambient environment, nanoscale adsorbed layers with different physicochemical properties can be sequentially removed. Using a sensor located a few nanometers beneath the surface, the characteristics of each layer can be precisely measured.
  (b) Adsorption analysis setup based on \NV center. The sample chamber is designed flow-through for vapor or liquid coming into contact with the diamond surface.
  } \label{setup}
\end{figure}

\begin{figure*}[ht]
  \begin{overpic}[width=0.95\textwidth]{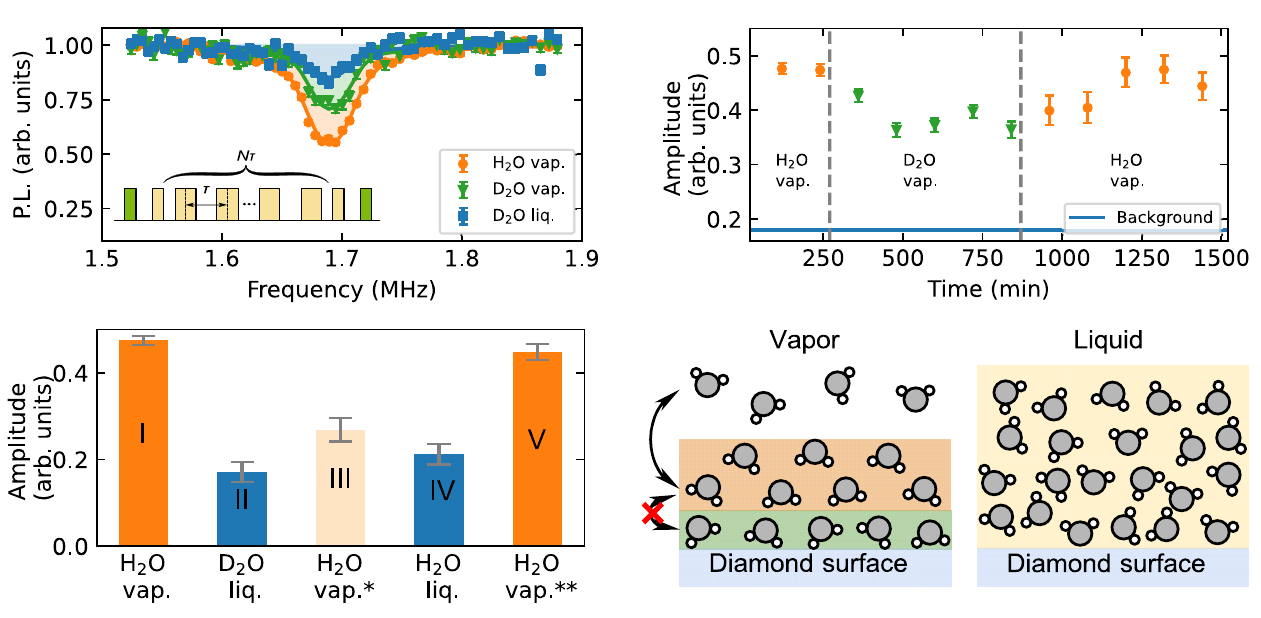} 
     \put (0, 47) {\bf a}
     \put (50, 47) {\bf b}
     \put (0, 23) {\bf c}
     \put (50, 23) {\bf d}
  \end{overpic}
  \caption[NV Center]{
  (a) Changing the environment allows for the replacement of isotopes in the water adsorption layer on the diamond surface. By altering the isotopes of the input gas, the isotopes of the water in the adsorption layer can be modified. Experimental proton \nmr spectra reveal differences in signal amplitudes for \hwater vapor (orange), \dwater vapor (green), and \dwater liquid (blue), demonstrating the influence of physical state and surface interactions on spectral dependence. An inset illustrates the NV-based \nmr control sequence (XY4-N), where the time interval $\tau$ in the control sequence determines the detection frequency at $1/2\tau$.
  (b) The resonant peak amplitudes in proton \nmr spectra exhibit temporal variations under different vapor environments. The orange and green data points in the figure correspond to measurements taken under \hwater and \dwater vapor conditions, respectively. The background signal (blue line), which was the surface state initialized by \dwater liquid, is used as the control in the study. The transition in vapor environments is indicated by the dashed line.
  (c) Comparison of the steady-state amplitudes of resonance peaks under changing environmental conditions involves resetting the surface to a proton surface (I) using \hwater liquid, replacing it with a deuterium surface (II) through the addition of \dwater liquid, and subsequently introducing \hwater vapor (III), \hwater liquid (IV), and \hwater vapor (V) in a sequence.
  (d) The schematic diagram depicting molecules on the diamond surface illustrates the distribution of the water layer on the diamond under vapor and liquid states. The left side represents vapor conditions, while the right side represents liquid conditions. It was observed that the low-dimensional water layer identified does not exchange with external water molecules under vapor conditions but can be activated under liquid conditions.} \label{vapor}
\end{figure*}

In this study, we developed a nanoscale liquid dissection method (Fig. \ref{setup}(a)) to analyze and differentiate surface water using NV center magnetic resonance technology. Water layers with diverse physical and chemical properties can be separated or eliminated through physical and chemical means. The capabilities of NV sensors in nanoscale magnetic resonance \cite{du2024Singlemolecule} enable precise analysis of nuclear magnetic resonance (\nmr) spectra and investigation of their structures, including the examination of interactions among different water layers (Fig. \ref{setup}(a)). Under an open ambient condition, an ice-like sub-nanometer water layer was observed on the diamond surface.  Remarkably, we observed that this water layer remained isolated from atmospheric water vapor and upper absorbed water layer throughout the experiment duration. Subsequently, we conducted a comprehensive examination and deliberation on this unexpected finding, leading to the formulation of a plausible explanatory model.


\begin{figure*}[hbtp]
  \begin{overpic}[width=0.95\textwidth]{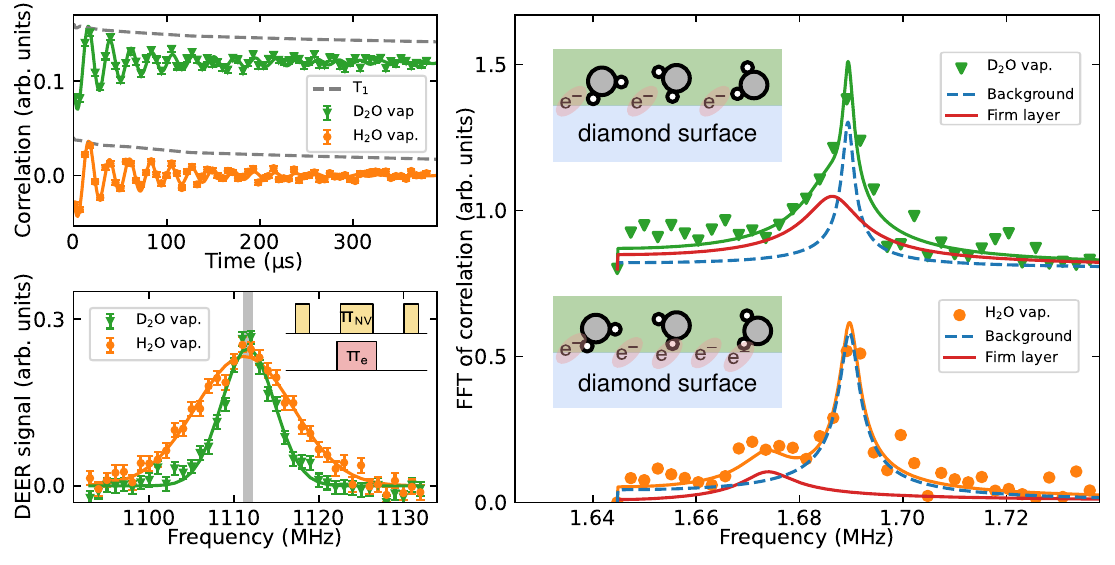} 
    \put (0, 51) {\bf a}
    \put (41, 51) {\bf b}
    \put (0, 26.5) {\bf c}
  \end{overpic}
  \caption[NV Center]
  {
  (a) Correlation signal obtained in \dwater vapor or \hwater vapor conditions. The sampling interval was set to 9.25 times the proton precession period. Corresponding \nv center relaxations are shown by gray dashed lines. 
  (b) The FFT amplitudes of the correlation signals shown in (a). The spectra reveal two primary frequency components of proton nuclei: a sharp peak corresponding to the slowly decaying signal at the proton precession frequency, and a broadened peak associated with the fast decaying signal in the firmly adsorbed layer, which shows a redshift relative to the proton precession frequency. The insets illustrate the mechanism of the redshift caused by paramagnetic shielding, with deuterium atoms highlighted in red.
  (c) DEER spectrum under \dwater vapor or \hwater vapor conditions. The diamond surface was \hwater liquid treated before experiment. The inset displays the pulse used in DEER experiments (laser pulses are not shown). The full width at half maximum of the DEER spectra is fitted to 7.2(3) MHz and 13.4(5) MHz, respectively, with a Gaussian lineshape.
  The upper bound of power broadening is indicated by a gray bar.
  }  \label{correlation}
\end{figure*}

We utilized a 50 $\mathrm{\mu m}$ thick ultrapure diamond film with a nitrogen concentration of 5 ppb as the substrate and fabricated near-surface NV centers as sensors via ion implantation technology \cite{du2024Singlemolecule}, with a depth of approximately 5-10 nm. The sensor was positioned inside a quartz enclosure, and the environmental conditions of the diamond surface are regulated by introducing gas or liquid into it, as illustrated in Fig. \ref{setup}(b). Using \odmr technique, the electron spin of NV centers is controlled via a coplanar microwave waveguide closely attached to the underside of the diamond. Concurrently, lasers are aimed at the shallow NV centers through the central aperture of the waveguide for the purpose of initializing and reading the NV centers.
The interfacial nuclear spins are weakly coupled to the shallow NV centers, therefore, \nmr spectra can be acquired via controls and readouts of the \nv center. For proton detection, we used dynamical decoupling sequences as shown in the inset of Fig. \ref{vapor}(a). The pulse sequence used was the standard XY4-N sequence, where NV spin flips were modulated by $\pi$ pulses at intervals of $\tau$ to detect the radio frequency signal at $\pi\tau$ frequency, corresponding to $\wl=\gH B_0$, with $\gH$ representing the gyromagnetic ratio of protons and $B_0$ denoting the static magnetic field applied.
Based on the \nmr spectra, information such as proton frequency, relaxation time, and quantity at the interface can be elucidated, as depicted in Fig. \ref{vapor}(a). 
Experimental studies were carried out under varying environmental conditions, covering \hwater vapor, \dwater vapor, and corresponding liquid environments. The near-saturated vapor was generated by passing ultrapure nitrogen ($> 99.9999\%$) through \hwater or \dwater water tank. By manipulating these conditions, the isotopes of the adsorbed layers were modified for distinct proton spectrum measurements. The proton spectrum showed the most prominent signal in the \hwater vapor environment, as illustrated in Fig. \ref{vapor}(a). Subsequent alteration of the \dwater vapor led to an exchange of water within the adsorption layer, resulting in a noticeable decrease in proton signal. Furthermore, the introduction of liquid \dwater further decreased the proton content, indicating the existence of intricate water layers with varying characteristics on the surface. Different substitutions occurred under different operational conditions, enabling a systematic exploration of the detailed spectral structures of complex water layers with diverse properties.

Firstly, we monitored the temporal evolution of proton spectral intensity as water vapor isotopes underwent exchange in the evolving environment (proton substituted for deuterium and vice versa), as depicted in Fig. \ref{vapor}(b). The diamond was initially immersed in a \hwater setting, dried, and subsequently exposed to a \hwater vapor atmosphere. A sufficient duration of measurement was taken to attain a stable state. After about 250 minutes, the environment was switched to \dwater vapor, resulting in a gradual decline in proton spectral intensity. Following a measurement duration of around 500 minutes, reintroduction of \hwater vapor led to gradual signal recovery in the proton nuclear magnetic spectrum (Fig. \ref{vapor}(b)). The transition from one stable state to another required approximately 3 hours. This suggests that a layer of adsorbed water undergoes gradual exchange with water molecules in the gas phase. 

Subsequently, the diamond was immersed in \dwater liquid, resulting in a further reduction in proton spectral intensity, yet not reaching zero (Phase II in Fig. \ref{vapor}(c)). The residual proton signals are likely attributed to various sources, including hydroxyl groups (-OH) present on the surface of oxidized diamond, as well as hydrogen-containing functional groups bound within subsurface defect regions (e.g., -CH$_2$ and -CH$3$). The introduction of \hwater vapor caused an increase in the proton spectral intensity (Phase III in Fig. \ref{vapor}(c)), although it remained significantly lower than the initial proton spectrum in the protonated surface state (Phase V in Fig. \ref{vapor}(c)).

These two experiments elucidate that certain adsorbed water molecules evade exchange with the gaseous environment even over a long experimental duration (record for a week). Consequently, two distinct layers of water adsorption are proposed (Fig. \ref{vapor}(d)): a loosely bound layer capable of atmospheric exchange and another layer activated solely upon substantial liquid water introduction. Through spectral amplitude analysis, it was determined that the thickness of this water layer is extremely thin, measuring less than 1 nm \cite{Supplementary}, thereby qualifying it as a \ldw layer. Moreover, an additional non-exchangeable proton layer, attributable to surface-residual hydrogenous functional groups, persists even upon liquid water introduction. This non-exchangeable hydrogen component, irrelevant to our study, will be factored out as background in subsequent analyses, with focus directed towards delineating the characteristics of the two distinct adsorbed water layers.

\begin{figure}[hbtp]
  \begin{overpic}[width=0.5\textwidth]{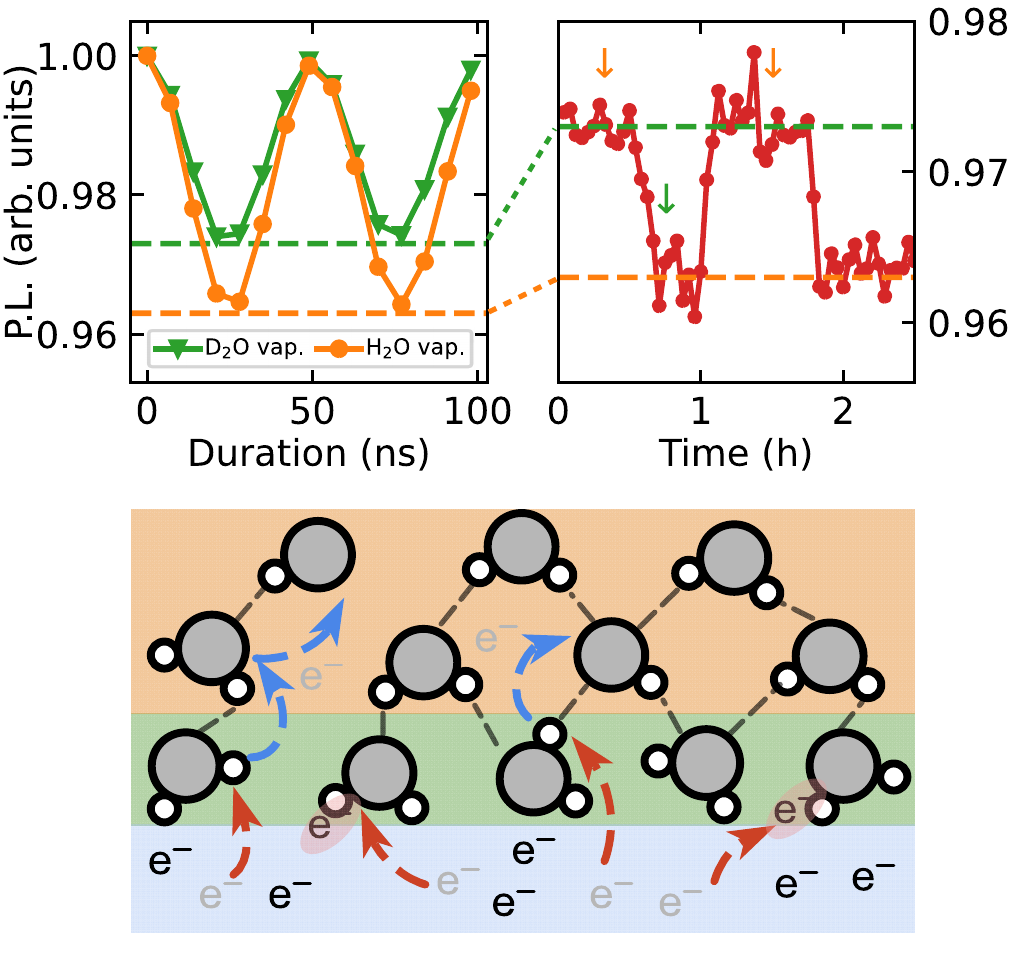} 
    \put (2, 91) {\bf a}
    \put (49, 91) {\bf b}
    \put (2, 44) {\bf c}
  \end{overpic}
  \caption[NV Center]{
  (a) Rabi oscillation of \nv center in \hwater  or \dwater vapor conditions.
  (b) The variation of photoluminescence difference between \nv center spin states during condition transitions. The orange (\hwater vapor) and green (\dwater vapor) arrows indicate when the corresponding transition begins.
  (c) Electron transfer model at diamond surface. The surface electrons are prone to transfer into the H bond network in the adsorption layer. The H bond energy affects the electron conductance in the network, resulting in different surface residual electron densities.
  }  \label{deer}
  \end{figure}
In order to investigate the impacts of the abnormal no-exchange layer further, we employed high-resolution \nmr techniques to analyze the intricate structure of proton spin on the diamond surface in both \hwater and \dwater vapor environments. Employing correlation spectroscopy methods \cite{staudacher2015Probing,kong2015Chemical} (Fig. \ref{correlation}(a)), we enhanced the spectral resolution from $\sim$MHz to the $\sim$kHz range. Fig. \ref{correlation}(b) exhibit the frequency domain \nmr spectra in the \hwater and \dwater vapor environments. In both cases, a distinct narrow spectral line is observed, which can be ascribed to surface proton-containing functional groups. Their spin-free neighboring environment or the rotational mobilities results in the reduction of interactions \cite{staudacher2015Probing,shi2015Singleprotein,shi2018SingleDNA}, leading to a narrow spectral width. In the \dwater vapor context of this comparative study, the upper loose layer was replaced with deuterium, causing its absence in the proton spectrum and a consequent reduction in the central peak intensity. Notably, we observed novel peaks deviating from the central peak, displaying a negative shift exceeding the spectrum linewidth in both \dwater and \hwater vapor environments, suggesting a potential chemical shift-like shielding effect. The observed frequency shift is 3 kHz in the \dwater vapor environment and 16 kHz in the \hwater vapor environment, displaying a roughly 1\% deviation at a magnetic field strength of $39.7\ \mathrm{ mT}$. This deviation indicates that there is stronger interaction between the \ldw layer and the  adsorbed water layer in the \hwater vapor environment, leading to a stronger shielding effect.

The redshift observed in nuclear spins could be attributed to electron shielding effects. To validate this, an investigation into the characteristics of surface electrons is essential. Fortunately, the NV center, functioning as a versatile quantum sensor, can employ \deer techniques to perform nanoscale \esr measurements with shallow NV centers \cite{shi2015Singleprotein}. Fig. \ref{correlation}(c)  demonstrates the \esr spectra obtained under \hwater and \dwater vapor conditions following the diamond surface pretreatment with \hwater liquid.  The \esr spectrum exhibits a noticeable broadening in the presence of \hwater vapor.  These observations imply a greater electron density  near the NV detector, offering additional evidence that the  electron  contributes to the redshift in nuclear spin resonance frequency from an alternative viewpoint.

The phenomenon is attributed to the localization of protons, which is known as nuclear quantum effects (NQE). NQE's presence results in an approximate 9\% increase in the number of hydrogen bonds in deuterated water compared to regular water \cite{flor2024Dissecting}, thereby boosting electron adsorption capacity and causing significant spectral changes in \ldw under both \hwater and \dwater vapor environments. This process diminishes the electron affinity of the diamond interface, as evidenced by the concurrent decrease in contrast of \nv centers (see Fig. \ref{deer} (a, b)). 
Transitioning to a \dwater vapor environment gradually diminishes the \nv center contrast, which can recover upon reverting to a \hwater vapor setting.
This reproducible effect underscores the heightened possibility of electron loss within illuminated NV centers under \hwater vapor conditions. It partially elucidates the potential electron sources, with potential contributions from other dark defects \cite{lozovoi2021Optical}, and the electron conductivity beyond the diamond \cite{xu2024Photoinduced}, as depicted in Fig. \ref{deer}(c). In \hwater settings, due to the NQE effect, more electrons may accumulate in the \ldw layer, enhancing the shielding effect on protons. This interpretation correlates with the observed variations in \nmr spectral chemical shifts and the broadening of \esr spectra.

In conclusion, our method, based on the \nv center, facilitates surface analysis to investigate the properties of interfacial water. By systematically substituting isotopes in various water layers, we distinguished different water layers and conducted separate analyses of the magnetic resonance spectrum, including nanoscale \nmr and \esr. Our research revealed the existence of an anomalous water layer, just one molecular layer thick, on the diamond surface at room temperature. This layer exhibits ice-like properties and remains stable without exchanging with external water vapor.High-resolution nanoscale \nmr analysis revealed significant chemical shifts of protons within this unique water layer, likely arising from the nuclear quantum effect of \dwater H bond. These findings provide valuable insights and experimental pathways for examining low-dimensional water structures and properties across the fields of materials science, geology, and biology.

\textbf{Acknowledgements:} 
We thank Prof. Sujing Wang and Prof. Jihu Su for helpful discussions. This work was supported by the National Natural Science Foundation of China (Grant No.\ T2125011), the CAS (Grants No.\ GJJSTD20200001, No.\ YSBR-068), Innovation Program for Quantum Science and Technology (Grants No.\ 2021ZD0302200,  No.\ 2021ZD0303204), New Cornerstone Science Foundation through the XPLORER PRIZE, and the Fundamental Research Funds for the Central Universities.

\textbf{Author contributions:} J.D. and F.S. supervised the project and proposed the idea; F.S., Z.L. and X.K. designed the experiments; Z.L. performed the experiments; H.S., P.Y. and Y.W. prepared the diamond sample; Z.L., X.K. and F.S. wrote the manuscript. All authors discussed and analyzed the data.
\bibliography{ref.bib}

\begin{thebibliography}{28}
\expandafter\ifx\csname natexlab\endcsname\relax\def\natexlab#1{#1}\fi
\expandafter\ifx\csname bibnamefont\endcsname\relax
  \def\bibnamefont#1{#1}\fi
\expandafter\ifx\csname bibfnamefont\endcsname\relax
  \def\bibfnamefont#1{#1}\fi
\expandafter\ifx\csname citenamefont\endcsname\relax
  \def\citenamefont#1{#1}\fi
\expandafter\ifx\csname url\endcsname\relax
  \def\url#1{\texttt{#1}}\fi
\expandafter\ifx\csname urlprefix\endcsname\relax\def\urlprefix{URL }\fi
\providecommand{\bibinfo}[2]{#2}
\providecommand{\eprint}[2][]{\url{#2}}

\bibitem[{\citenamefont{Salzmann}(2019)}]{salzmann2019Advances}
\bibinfo{author}{\bibfnamefont{C.~G.} \bibnamefont{Salzmann}}, \bibinfo{journal}{The Journal of Chemical Physics} \textbf{\bibinfo{volume}{150}}, \bibinfo{pages}{060901} (\bibinfo{year}{2019}).

\bibitem[{\citenamefont{Ball}(2008)}]{ball2008Water}
\bibinfo{author}{\bibfnamefont{P.}~\bibnamefont{Ball}}, \bibinfo{journal}{Chem. Rev.} \textbf{\bibinfo{volume}{108}}, \bibinfo{pages}{74} (\bibinfo{year}{2008}).

\bibitem[{\citenamefont{Tanford}(1962)}]{tanford1962Contribution}
\bibinfo{author}{\bibfnamefont{{\relax Charles}.}~\bibnamefont{Tanford}}, \bibinfo{journal}{J. Am. Chem. Soc.} \textbf{\bibinfo{volume}{84}}, \bibinfo{pages}{4240} (\bibinfo{year}{1962}).

\bibitem[{\citenamefont{Wu et~al.}(2017)\citenamefont{Wu, Chen, Li, Li, Xu, and Dong}}]{wu2017Wettability}
\bibinfo{author}{\bibfnamefont{K.}~\bibnamefont{Wu}}, \bibinfo{author}{\bibfnamefont{Z.}~\bibnamefont{Chen}}, \bibinfo{author}{\bibfnamefont{J.}~\bibnamefont{Li}}, \bibinfo{author}{\bibfnamefont{X.}~\bibnamefont{Li}}, \bibinfo{author}{\bibfnamefont{J.}~\bibnamefont{Xu}}, \bibnamefont{and} \bibinfo{author}{\bibfnamefont{X.}~\bibnamefont{Dong}}, \bibinfo{journal}{Proc. Natl. Acad. Sci.} \textbf{\bibinfo{volume}{114}}, \bibinfo{pages}{3358} (\bibinfo{year}{2017}).

\bibitem[{\citenamefont{Davies et~al.}(2024)\citenamefont{Davies, {Rosu-Finsen}, Salzmann, and Michaelides}}]{davies2024How}
\bibinfo{author}{\bibfnamefont{M.~B.} \bibnamefont{Davies}}, \bibinfo{author}{\bibfnamefont{A.}~\bibnamefont{{Rosu-Finsen}}}, \bibinfo{author}{\bibfnamefont{C.~G.} \bibnamefont{Salzmann}}, \bibnamefont{and} \bibinfo{author}{\bibfnamefont{A.}~\bibnamefont{Michaelides}}, \emph{\bibinfo{title}{How {{Crystalline}} is {{Low-Density Amorphous Ice}}?}} (\bibinfo{year}{2024}), \eprint{2305.03057}.

\bibitem[{\citenamefont{Kapil et~al.}(2022)\citenamefont{Kapil, Schran, Zen, Chen, Pickard, and Michaelides}}]{kapil2022Firstprinciples}
\bibinfo{author}{\bibfnamefont{V.}~\bibnamefont{Kapil}}, \bibinfo{author}{\bibfnamefont{C.}~\bibnamefont{Schran}}, \bibinfo{author}{\bibfnamefont{A.}~\bibnamefont{Zen}}, \bibinfo{author}{\bibfnamefont{J.}~\bibnamefont{Chen}}, \bibinfo{author}{\bibfnamefont{C.~J.} \bibnamefont{Pickard}}, \bibnamefont{and} \bibinfo{author}{\bibfnamefont{A.}~\bibnamefont{Michaelides}}, \bibinfo{journal}{Nature} \textbf{\bibinfo{volume}{609}}, \bibinfo{pages}{512} (\bibinfo{year}{2022}).

\bibitem[{\citenamefont{Nie et~al.}(2010)\citenamefont{Nie, Feibelman, Bartelt, and Th{\"u}rmer}}]{nie2010Pentagons}
\bibinfo{author}{\bibfnamefont{S.}~\bibnamefont{Nie}}, \bibinfo{author}{\bibfnamefont{P.~J.} \bibnamefont{Feibelman}}, \bibinfo{author}{\bibfnamefont{N.~C.} \bibnamefont{Bartelt}}, \bibnamefont{and} \bibinfo{author}{\bibfnamefont{K.}~\bibnamefont{Th{\"u}rmer}}, \bibinfo{journal}{Phys. Rev. Lett.} \textbf{\bibinfo{volume}{105}}, \bibinfo{pages}{026102} (\bibinfo{year}{2010}).

\bibitem[{\citenamefont{Maier et~al.}(2016)\citenamefont{Maier, Lechner, Somorjai, and Salmeron}}]{maier2016Growth}
\bibinfo{author}{\bibfnamefont{S.}~\bibnamefont{Maier}}, \bibinfo{author}{\bibfnamefont{B.~A.~J.} \bibnamefont{Lechner}}, \bibinfo{author}{\bibfnamefont{G.~A.} \bibnamefont{Somorjai}}, \bibnamefont{and} \bibinfo{author}{\bibfnamefont{M.}~\bibnamefont{Salmeron}}, \bibinfo{journal}{J. Am. Chem. Soc.} \textbf{\bibinfo{volume}{138}}, \bibinfo{pages}{3145} (\bibinfo{year}{2016}).

\bibitem[{\citenamefont{Lin et~al.}(2018)\citenamefont{Lin, Avidor, Corem, Godsi, Alexandrowicz, Darling, and Hodgson}}]{lin2018TwoDimensional}
\bibinfo{author}{\bibfnamefont{C.}~\bibnamefont{Lin}}, \bibinfo{author}{\bibfnamefont{N.}~\bibnamefont{Avidor}}, \bibinfo{author}{\bibfnamefont{G.}~\bibnamefont{Corem}}, \bibinfo{author}{\bibfnamefont{O.}~\bibnamefont{Godsi}}, \bibinfo{author}{\bibfnamefont{G.}~\bibnamefont{Alexandrowicz}}, \bibinfo{author}{\bibfnamefont{G.~R.} \bibnamefont{Darling}}, \bibnamefont{and} \bibinfo{author}{\bibfnamefont{A.}~\bibnamefont{Hodgson}}, \bibinfo{journal}{Phys. Rev. Lett.} \textbf{\bibinfo{volume}{120}}, \bibinfo{pages}{076101} (\bibinfo{year}{2018}).

\bibitem[{\citenamefont{Girit et~al.}(2009)\citenamefont{Girit, Meyer, Erni, Rossell, Kisielowski, Yang, Park, Crommie, Cohen, Louie et~al.}}]{girit2009Graphene}
\bibinfo{author}{\bibfnamefont{{\c C}.~{\"O}.} \bibnamefont{Girit}}, \bibinfo{author}{\bibfnamefont{J.~C.} \bibnamefont{Meyer}}, \bibinfo{author}{\bibfnamefont{R.}~\bibnamefont{Erni}}, \bibinfo{author}{\bibfnamefont{M.~D.} \bibnamefont{Rossell}}, \bibinfo{author}{\bibfnamefont{C.}~\bibnamefont{Kisielowski}}, \bibinfo{author}{\bibfnamefont{L.}~\bibnamefont{Yang}}, \bibinfo{author}{\bibfnamefont{C.-H.} \bibnamefont{Park}}, \bibinfo{author}{\bibfnamefont{M.~F.} \bibnamefont{Crommie}}, \bibinfo{author}{\bibfnamefont{M.~L.} \bibnamefont{Cohen}}, \bibinfo{author}{\bibfnamefont{S.~G.} \bibnamefont{Louie}}, \bibnamefont{et~al.}, \bibinfo{journal}{Science} \textbf{\bibinfo{volume}{323}}, \bibinfo{pages}{1705} (\bibinfo{year}{2009}).

\bibitem[{\citenamefont{{Algara-Siller} et~al.}(2015)\citenamefont{{Algara-Siller}, Lehtinen, Wang, Nair, Kaiser, Wu, Geim, and Grigorieva}}]{algara-siller2015Square}
\bibinfo{author}{\bibfnamefont{G.}~\bibnamefont{{Algara-Siller}}}, \bibinfo{author}{\bibfnamefont{O.}~\bibnamefont{Lehtinen}}, \bibinfo{author}{\bibfnamefont{F.~C.} \bibnamefont{Wang}}, \bibinfo{author}{\bibfnamefont{R.~R.} \bibnamefont{Nair}}, \bibinfo{author}{\bibfnamefont{U.}~\bibnamefont{Kaiser}}, \bibinfo{author}{\bibfnamefont{H.~A.} \bibnamefont{Wu}}, \bibinfo{author}{\bibfnamefont{A.~K.} \bibnamefont{Geim}}, \bibnamefont{and} \bibinfo{author}{\bibfnamefont{I.~V.} \bibnamefont{Grigorieva}}, \bibinfo{journal}{Nature} \textbf{\bibinfo{volume}{519}}, \bibinfo{pages}{443} (\bibinfo{year}{2015}).

\bibitem[{\citenamefont{Peng et~al.}(2018)\citenamefont{Peng, Cao, He, Guo, Hapala, Ma, Cheng, Chen, Xie, Li et~al.}}]{peng2018Effect}
\bibinfo{author}{\bibfnamefont{J.}~\bibnamefont{Peng}}, \bibinfo{author}{\bibfnamefont{D.}~\bibnamefont{Cao}}, \bibinfo{author}{\bibfnamefont{Z.}~\bibnamefont{He}}, \bibinfo{author}{\bibfnamefont{J.}~\bibnamefont{Guo}}, \bibinfo{author}{\bibfnamefont{P.}~\bibnamefont{Hapala}}, \bibinfo{author}{\bibfnamefont{R.}~\bibnamefont{Ma}}, \bibinfo{author}{\bibfnamefont{B.}~\bibnamefont{Cheng}}, \bibinfo{author}{\bibfnamefont{J.}~\bibnamefont{Chen}}, \bibinfo{author}{\bibfnamefont{W.~J.} \bibnamefont{Xie}}, \bibinfo{author}{\bibfnamefont{X.-Z.} \bibnamefont{Li}}, \bibnamefont{et~al.}, \bibinfo{journal}{Nature} \textbf{\bibinfo{volume}{557}}, \bibinfo{pages}{701} (\bibinfo{year}{2018}).

\bibitem[{\citenamefont{Shiotari and Sugimoto}(2017)}]{shiotari2017Ultrahighresolution}
\bibinfo{author}{\bibfnamefont{A.}~\bibnamefont{Shiotari}} \bibnamefont{and} \bibinfo{author}{\bibfnamefont{Y.}~\bibnamefont{Sugimoto}}, \bibinfo{journal}{Nat Commun} \textbf{\bibinfo{volume}{8}}, \bibinfo{pages}{14313} (\bibinfo{year}{2017}).

\bibitem[{\citenamefont{Ma et~al.}(2020)\citenamefont{Ma, Cao, Zhu, Tian, Peng, Guo, Chen, Li, Francisco, Zeng et~al.}}]{ma2020Atomic}
\bibinfo{author}{\bibfnamefont{R.}~\bibnamefont{Ma}}, \bibinfo{author}{\bibfnamefont{D.}~\bibnamefont{Cao}}, \bibinfo{author}{\bibfnamefont{C.}~\bibnamefont{Zhu}}, \bibinfo{author}{\bibfnamefont{Y.}~\bibnamefont{Tian}}, \bibinfo{author}{\bibfnamefont{J.}~\bibnamefont{Peng}}, \bibinfo{author}{\bibfnamefont{J.}~\bibnamefont{Guo}}, \bibinfo{author}{\bibfnamefont{J.}~\bibnamefont{Chen}}, \bibinfo{author}{\bibfnamefont{X.-Z.} \bibnamefont{Li}}, \bibinfo{author}{\bibfnamefont{J.~S.} \bibnamefont{Francisco}}, \bibinfo{author}{\bibfnamefont{X.~C.} \bibnamefont{Zeng}}, \bibnamefont{et~al.}, \bibinfo{journal}{Nature} \textbf{\bibinfo{volume}{577}}, \bibinfo{pages}{60} (\bibinfo{year}{2020}).

\bibitem[{\citenamefont{Du et~al.}(2024)\citenamefont{Du, Shi, Kong, Jelezko, and Wrachtrup}}]{du2024Singlemolecule}
\bibinfo{author}{\bibfnamefont{J.}~\bibnamefont{Du}}, \bibinfo{author}{\bibfnamefont{F.}~\bibnamefont{Shi}}, \bibinfo{author}{\bibfnamefont{X.}~\bibnamefont{Kong}}, \bibinfo{author}{\bibfnamefont{F.}~\bibnamefont{Jelezko}}, \bibnamefont{and} \bibinfo{author}{\bibfnamefont{J.}~\bibnamefont{Wrachtrup}}, \bibinfo{journal}{Rev. Mod. Phys.} \textbf{\bibinfo{volume}{96}}, \bibinfo{pages}{025001} (\bibinfo{year}{2024}).

\bibitem[{\citenamefont{Claridge et~al.}(2011)\citenamefont{Claridge, Schwartz, and Weiss}}]{claridge2011Electrons}
\bibinfo{author}{\bibfnamefont{S.~A.} \bibnamefont{Claridge}}, \bibinfo{author}{\bibfnamefont{J.~J.} \bibnamefont{Schwartz}}, \bibnamefont{and} \bibinfo{author}{\bibfnamefont{P.~S.} \bibnamefont{Weiss}}, \bibinfo{journal}{ACS Nano} \textbf{\bibinfo{volume}{5}}, \bibinfo{pages}{693} (\bibinfo{year}{2011}).

\bibitem[{\citenamefont{Wang et~al.}(2019)\citenamefont{Wang, Chen, Guo, Peng, Wang, Chen, Ma, Zhang, Su, Rong et~al.}}]{wang2019Nanoscale}
\bibinfo{author}{\bibfnamefont{P.}~\bibnamefont{Wang}}, \bibinfo{author}{\bibfnamefont{S.}~\bibnamefont{Chen}}, \bibinfo{author}{\bibfnamefont{M.}~\bibnamefont{Guo}}, \bibinfo{author}{\bibfnamefont{S.}~\bibnamefont{Peng}}, \bibinfo{author}{\bibfnamefont{M.}~\bibnamefont{Wang}}, \bibinfo{author}{\bibfnamefont{M.}~\bibnamefont{Chen}}, \bibinfo{author}{\bibfnamefont{W.}~\bibnamefont{Ma}}, \bibinfo{author}{\bibfnamefont{R.}~\bibnamefont{Zhang}}, \bibinfo{author}{\bibfnamefont{J.}~\bibnamefont{Su}}, \bibinfo{author}{\bibfnamefont{X.}~\bibnamefont{Rong}}, \bibnamefont{et~al.}, \bibinfo{journal}{Sci. Adv.} \textbf{\bibinfo{volume}{5}}, \bibinfo{pages}{eaau8038} (\bibinfo{year}{2019}).

\bibitem[{\citenamefont{Grinolds et~al.}(2014)\citenamefont{Grinolds, Warner, Greve, Dovzhenko, Thiel, Walsworth, Hong, Maletinsky, and Yacoby}}]{grinolds2014Subnanometre}
\bibinfo{author}{\bibfnamefont{M.~S.} \bibnamefont{Grinolds}}, \bibinfo{author}{\bibfnamefont{M.}~\bibnamefont{Warner}}, \bibinfo{author}{\bibfnamefont{K.~D.} \bibnamefont{Greve}}, \bibinfo{author}{\bibfnamefont{Y.}~\bibnamefont{Dovzhenko}}, \bibinfo{author}{\bibfnamefont{L.}~\bibnamefont{Thiel}}, \bibinfo{author}{\bibfnamefont{R.~L.} \bibnamefont{Walsworth}}, \bibinfo{author}{\bibfnamefont{S.}~\bibnamefont{Hong}}, \bibinfo{author}{\bibfnamefont{P.}~\bibnamefont{Maletinsky}}, \bibnamefont{and} \bibinfo{author}{\bibfnamefont{A.}~\bibnamefont{Yacoby}}, \bibinfo{journal}{Nat Nano} \textbf{\bibinfo{volume}{9}}, \bibinfo{pages}{279} (\bibinfo{year}{2014}).

\bibitem[{\citenamefont{Abobeih et~al.}(2019)\citenamefont{Abobeih, Randall, Bradley, Bartling, Bakker, Degen, Markham, Twitchen, and Taminiau}}]{abobeih2019Atomicscale}
\bibinfo{author}{\bibfnamefont{M.~H.} \bibnamefont{Abobeih}}, \bibinfo{author}{\bibfnamefont{J.}~\bibnamefont{Randall}}, \bibinfo{author}{\bibfnamefont{C.~E.} \bibnamefont{Bradley}}, \bibinfo{author}{\bibfnamefont{H.~P.} \bibnamefont{Bartling}}, \bibinfo{author}{\bibfnamefont{M.~A.} \bibnamefont{Bakker}}, \bibinfo{author}{\bibfnamefont{M.~J.} \bibnamefont{Degen}}, \bibinfo{author}{\bibfnamefont{M.}~\bibnamefont{Markham}}, \bibinfo{author}{\bibfnamefont{D.~J.} \bibnamefont{Twitchen}}, \bibnamefont{and} \bibinfo{author}{\bibfnamefont{T.~H.} \bibnamefont{Taminiau}}, \bibinfo{journal}{Nature} \textbf{\bibinfo{volume}{576}}, \bibinfo{pages}{411} (\bibinfo{year}{2019}).

\bibitem[{\citenamefont{Aslam et~al.}(2017)\citenamefont{Aslam, Pfender, Neumann, Reuter, Zappe, de~Oliveira, Denisenko, Sumiya, Onoda, Isoya et~al.}}]{aslam2017Nanoscale}
\bibinfo{author}{\bibfnamefont{N.}~\bibnamefont{Aslam}}, \bibinfo{author}{\bibfnamefont{M.}~\bibnamefont{Pfender}}, \bibinfo{author}{\bibfnamefont{P.}~\bibnamefont{Neumann}}, \bibinfo{author}{\bibfnamefont{R.}~\bibnamefont{Reuter}}, \bibinfo{author}{\bibfnamefont{A.}~\bibnamefont{Zappe}}, \bibinfo{author}{\bibfnamefont{F.~F.} \bibnamefont{de~Oliveira}}, \bibinfo{author}{\bibfnamefont{A.}~\bibnamefont{Denisenko}}, \bibinfo{author}{\bibfnamefont{H.}~\bibnamefont{Sumiya}}, \bibinfo{author}{\bibfnamefont{S.}~\bibnamefont{Onoda}}, \bibinfo{author}{\bibfnamefont{J.}~\bibnamefont{Isoya}}, \bibnamefont{et~al.}, \bibinfo{journal}{Science} \textbf{\bibinfo{volume}{357}}, \bibinfo{pages}{67} (\bibinfo{year}{2017}).

\bibitem[{Sup()}]{Supplementary}
\emph{\bibinfo{title}{Supplementary {{Material}}}}.

\bibitem[{\citenamefont{Staudacher et~al.}(2015)\citenamefont{Staudacher, Raatz, Pezzagna, Meijer, Reinhard, Meriles, and Wrachtrup}}]{staudacher2015Probing}
\bibinfo{author}{\bibfnamefont{T.}~\bibnamefont{Staudacher}}, \bibinfo{author}{\bibfnamefont{N.}~\bibnamefont{Raatz}}, \bibinfo{author}{\bibfnamefont{S.}~\bibnamefont{Pezzagna}}, \bibinfo{author}{\bibfnamefont{J.}~\bibnamefont{Meijer}}, \bibinfo{author}{\bibfnamefont{F.}~\bibnamefont{Reinhard}}, \bibinfo{author}{\bibfnamefont{C.~A.} \bibnamefont{Meriles}}, \bibnamefont{and} \bibinfo{author}{\bibfnamefont{J.}~\bibnamefont{Wrachtrup}}, \bibinfo{journal}{Nat. Commun.} \textbf{\bibinfo{volume}{6}}, \bibinfo{pages}{8527} (\bibinfo{year}{2015}).

\bibitem[{\citenamefont{Kong et~al.}(2015)\citenamefont{Kong, Stark, Du, McGuinness, and Jelezko}}]{kong2015Chemical}
\bibinfo{author}{\bibfnamefont{X.}~\bibnamefont{Kong}}, \bibinfo{author}{\bibfnamefont{A.}~\bibnamefont{Stark}}, \bibinfo{author}{\bibfnamefont{J.}~\bibnamefont{Du}}, \bibinfo{author}{\bibfnamefont{L.~P.} \bibnamefont{McGuinness}}, \bibnamefont{and} \bibinfo{author}{\bibfnamefont{F.}~\bibnamefont{Jelezko}}, \bibinfo{journal}{Phys. Rev. Applied} \textbf{\bibinfo{volume}{4}}, \bibinfo{pages}{024004} (\bibinfo{year}{2015}).

\bibitem[{\citenamefont{Shi et~al.}(2015)\citenamefont{Shi, Zhang, Wang, Sun, Wang, Rong, Chen, Ju, Reinhard, Chen et~al.}}]{shi2015Singleprotein}
\bibinfo{author}{\bibfnamefont{F.}~\bibnamefont{Shi}}, \bibinfo{author}{\bibfnamefont{Q.}~\bibnamefont{Zhang}}, \bibinfo{author}{\bibfnamefont{P.}~\bibnamefont{Wang}}, \bibinfo{author}{\bibfnamefont{H.}~\bibnamefont{Sun}}, \bibinfo{author}{\bibfnamefont{J.}~\bibnamefont{Wang}}, \bibinfo{author}{\bibfnamefont{X.}~\bibnamefont{Rong}}, \bibinfo{author}{\bibfnamefont{M.}~\bibnamefont{Chen}}, \bibinfo{author}{\bibfnamefont{C.}~\bibnamefont{Ju}}, \bibinfo{author}{\bibfnamefont{F.}~\bibnamefont{Reinhard}}, \bibinfo{author}{\bibfnamefont{H.}~\bibnamefont{Chen}}, \bibnamefont{et~al.}, \bibinfo{journal}{Science} \textbf{\bibinfo{volume}{347}}, \bibinfo{pages}{1135} (\bibinfo{year}{2015}).

\bibitem[{\citenamefont{Shi et~al.}(2018)\citenamefont{Shi, Kong, Zhao, Zhang, Chen, Chen, Zhang, Wang, Ye, Wang et~al.}}]{shi2018SingleDNA}
\bibinfo{author}{\bibfnamefont{F.}~\bibnamefont{Shi}}, \bibinfo{author}{\bibfnamefont{F.}~\bibnamefont{Kong}}, \bibinfo{author}{\bibfnamefont{P.}~\bibnamefont{Zhao}}, \bibinfo{author}{\bibfnamefont{X.}~\bibnamefont{Zhang}}, \bibinfo{author}{\bibfnamefont{M.}~\bibnamefont{Chen}}, \bibinfo{author}{\bibfnamefont{S.}~\bibnamefont{Chen}}, \bibinfo{author}{\bibfnamefont{Q.}~\bibnamefont{Zhang}}, \bibinfo{author}{\bibfnamefont{M.}~\bibnamefont{Wang}}, \bibinfo{author}{\bibfnamefont{X.}~\bibnamefont{Ye}}, \bibinfo{author}{\bibfnamefont{Z.}~\bibnamefont{Wang}}, \bibnamefont{et~al.}, \bibinfo{journal}{Nat. Methods} \textbf{\bibinfo{volume}{15}}, \bibinfo{pages}{697} (\bibinfo{year}{2018}).

\bibitem[{\citenamefont{Fl{\'o}r et~al.}(2024)\citenamefont{Fl{\'o}r, Wilkins, {de la Puente}, Laage, Cassone, Hassanali, and Roke}}]{flor2024Dissecting}
\bibinfo{author}{\bibfnamefont{M.}~\bibnamefont{Fl{\'o}r}}, \bibinfo{author}{\bibfnamefont{D.~M.} \bibnamefont{Wilkins}}, \bibinfo{author}{\bibfnamefont{M.}~\bibnamefont{{de la Puente}}}, \bibinfo{author}{\bibfnamefont{D.}~\bibnamefont{Laage}}, \bibinfo{author}{\bibfnamefont{G.}~\bibnamefont{Cassone}}, \bibinfo{author}{\bibfnamefont{A.}~\bibnamefont{Hassanali}}, \bibnamefont{and} \bibinfo{author}{\bibfnamefont{S.}~\bibnamefont{Roke}}, \bibinfo{journal}{Science} \textbf{\bibinfo{volume}{386}}, \bibinfo{pages}{eads4369} (\bibinfo{year}{2024}).

\bibitem[{\citenamefont{Lozovoi et~al.}(2021)\citenamefont{Lozovoi, Jayakumar, Daw, Vizkelethy, Bielejec, Doherty, Flick, and Meriles}}]{lozovoi2021Optical}
\bibinfo{author}{\bibfnamefont{A.}~\bibnamefont{Lozovoi}}, \bibinfo{author}{\bibfnamefont{H.}~\bibnamefont{Jayakumar}}, \bibinfo{author}{\bibfnamefont{D.}~\bibnamefont{Daw}}, \bibinfo{author}{\bibfnamefont{G.}~\bibnamefont{Vizkelethy}}, \bibinfo{author}{\bibfnamefont{E.}~\bibnamefont{Bielejec}}, \bibinfo{author}{\bibfnamefont{M.~W.} \bibnamefont{Doherty}}, \bibinfo{author}{\bibfnamefont{J.}~\bibnamefont{Flick}}, \bibnamefont{and} \bibinfo{author}{\bibfnamefont{C.~A.} \bibnamefont{Meriles}}, \bibinfo{journal}{Nat Electron} \textbf{\bibinfo{volume}{4}}, \bibinfo{pages}{717} (\bibinfo{year}{2021}).

\bibitem[{\citenamefont{Xu et~al.}(2024)\citenamefont{Xu, Pagliero, {L{\'o}pez-Morales}, Flick, Wolcott, and Meriles}}]{xu2024Photoinduced}
\bibinfo{author}{\bibfnamefont{K.}~\bibnamefont{Xu}}, \bibinfo{author}{\bibfnamefont{D.}~\bibnamefont{Pagliero}}, \bibinfo{author}{\bibfnamefont{G.~I.} \bibnamefont{{L{\'o}pez-Morales}}}, \bibinfo{author}{\bibfnamefont{J.}~\bibnamefont{Flick}}, \bibinfo{author}{\bibfnamefont{A.}~\bibnamefont{Wolcott}}, \bibnamefont{and} \bibinfo{author}{\bibfnamefont{C.~A.} \bibnamefont{Meriles}}, \bibinfo{journal}{ACS Appl. Mater. Interfaces} \textbf{\bibinfo{volume}{16}}, \bibinfo{pages}{37226} (\bibinfo{year}{2024}).

\end{thebibliography}
\end{document}